\documentclass[journal]{IEEEtran}

\ifCLASSINFOpdf
\else
   \usepackage[dvips]{graphicx}
\fi
\usepackage{url}

\usepackage{graphicx}
\usepackage{amsmath,bm,graphicx}
\usepackage{algorithmic}
\usepackage{mathtools}
\usepackage{multirow}
\usepackage{enumerate,enumitem}
\usepackage{graphicx}
\usepackage{subcaption}
\usepackage[ruled,vlined]{algorithm2e}
\usepackage{xcolor}
\usepackage{cite}
\newcommand{\trans}{^{\mathsf{T}}}
\newcommand{\sigmaue}{\frac{1}{d}\mathbf{\Sigma}_\text{Ue}}
\newcommand{\sigmaut}{\frac{1}{d}\mathbf{\Sigma}_\text{Ut}}
\DeclarePairedDelimiter\norm{\lVert}{\rVert}

\begin{document}

\title{Cosine Scoring with Uncertainty for Neural Speaker Embedding}

\author{Qiongqiong Wang, Kong Aik Lee,~\IEEEmembership{Senior Member,~IEEE}
\thanks{Qiongqiong Wang is with the Institute for Infocomm Research (I$^2$R), Agency for Science, Technology and Research (A$^\star$STAR), Singapore 138632  (e-mail: wang\underline{\enskip}qiongqiong@i2r.a-star.edu.sg).}
\thanks{Kong Aik Lee is with the Dept. of Electrical and Electronic Engineering, The Hong Kong Polytechnic University, Hung Hom, Kowloon, Hong Kong (email: kong-aik.lee@polyu.edu.hk)}}


\maketitle
\begin{abstract}
Uncertainty modeling in speaker representation aims to learn the variability present in speech utterances. While the conventional cosine-scoring is computationally efficient and prevalent in speaker recognition, it lacks the capability to handle uncertainty. 
To address this challenge, this paper proposes an approach for estimating uncertainty at the speaker embedding front-end and propagating it to the cosine scoring back-end.
Experiments conducted on the VoxCeleb and SITW datasets confirmed the efficacy of the proposed method in handling uncertainty arising from embedding estimation. It achieved improvement with 8.5\% and 9.8\% average reductions in EER and minDCF compared to the conventional cosine similarity. It is also computationally efficient in practice.  
\end{abstract}


\IEEEpeerreviewmaketitle

\section{Introduction}
\label{sec:intro}
\IEEEPARstart{S}{peaker} recognition is the automated process of recognizing who is speaking by analyzing the speech and extracting speaker-specific information~\cite{bai21}. 
Modern speaker recognition systems consist of a front-end that produces speaker embeddings followed by a scoring back-end~\cite{Matejka20,villalba20,nagrani20, lee20, Ferrera22}. Speaker embeddings are fixed-length continuous-value representations of input sequences of acoustic feature vectors~\cite{Brummer18}. Representing speech utterances as embedding simplifies speaker recognition tasks because simple geometric operations are easily applicable in the embedding space. 

A speaker-embedding neural network consists of three main modules~\cite{lee21}: (1) a frame-level feature encoder, (2) a temporal aggregation pooling layer that converts frame-level features into a fixed-dimensional condensed vector, and (3) a decoder that produces the embedding representations. During training, a softmax function is used at the output layer, and a cross-entropy loss is used for network optimization. 
A substantial amount of work has been reported on neural network architectures that produce embedding vectors with improved speaker representations~\cite{lee21, Snyder18, Okabe18, Chung18, Rohdin20, liu22,liu2023golden}. Commonly used scoring back-ends include the \textit{probabilistic linear discriminant analysis} (PLDA)~\cite{ioffe06,prince07, wang23Generalized} and \textit{cosine similarity} measure. 

Speech utterances exhibit intrinsic variability owing to the physiological nature of the vocal apparatus and psychological states such as emotion and mental health, as well as extrinsic variability from background noises, channel distortions, and duration~\cite{lee21}. Despite being speaker-independent, the extrinsic variabilities do affect how representative the extracted embedding vectors are of the speaker's voice~\cite{takahashi14}. For example, it has been noted that speech utterances of shorter length generate less dependable embeddings, leading to poor performance in speaker verification~\cite{Zeinali20}.
Embeddings of the same utterance extracted using different networks may also differ in terms of their representative power. It depends on the conditions and amount of training data, the network capacity, and so on. 
Therefore, measuring the precision, or equivalently, the uncertainty of the embeddings is essential.
 
Considering uncertainty in both front-end embedding extraction and back-end scoring has been found effective. For the front-end, an auxiliary neural network has been employed to predict the uncertainty by utilizing the output from the statistics pooling layer of the original x-vector~\cite{Silnova20}. A posterior inference pooling has been proposed in a speaker-embedding neural network, referred to as the xi-vector network~\cite{lee21,liu2023disentangling}. It predicts the uncertainty of each frame in the input and subsequently propagates it to the embedding estimation. In~\cite{qwang23}, an uncertainty-propagated PLDA (UP-PLDA) introduces an increase in the within-speaker covariance by the posterior covariance corresponding to the embedding uncertainty. It improves the robustness against local and random perturbation, an inherent property of speech utterances.

In addition to the speaker-embedding network architectures, advances in loss functions have brought significant progress in speaker representation learning~\cite{Wan18, Liu19, Zhou20, Li22}. 
To increase the compactness within the same speaker class, various techniques incorporate margin penalties to the conventional softmax activation function following a cross-entropy loss~\cite{hwang18,fwang18,deng18}.  The use of large-margin embeddings amplifies speaker-discriminative capabilities, leading to a shift from the PLDA back-end to a more straightforward cosine similarity measure.  While cosine scoring is computationally efficient and does not necessitate training, it lacks the capability to manage uncertainty~\cite{Liu19, Zhou20, Desplanques20, wang22}.

This paper explores methods for uncertainty propagation from the embeddings to the back-end of speaker recognition tasks, specifically, cosine scoring. To the best of our knowledge, this is the first work taking uncertainty into consideration for cosine scoring.
\vspace{-0.2cm}
\section{Speaker embeddings and cosine scoring}
The estimation of the speaker embedding is influenced by the variability in utterances and the property of embedding extraction networks. Hence, it is essential to measure the uncertainty, or precision, of the embeddings.

\begin{figure}[t]
\centering
\includegraphics[width=0.75\columnwidth]{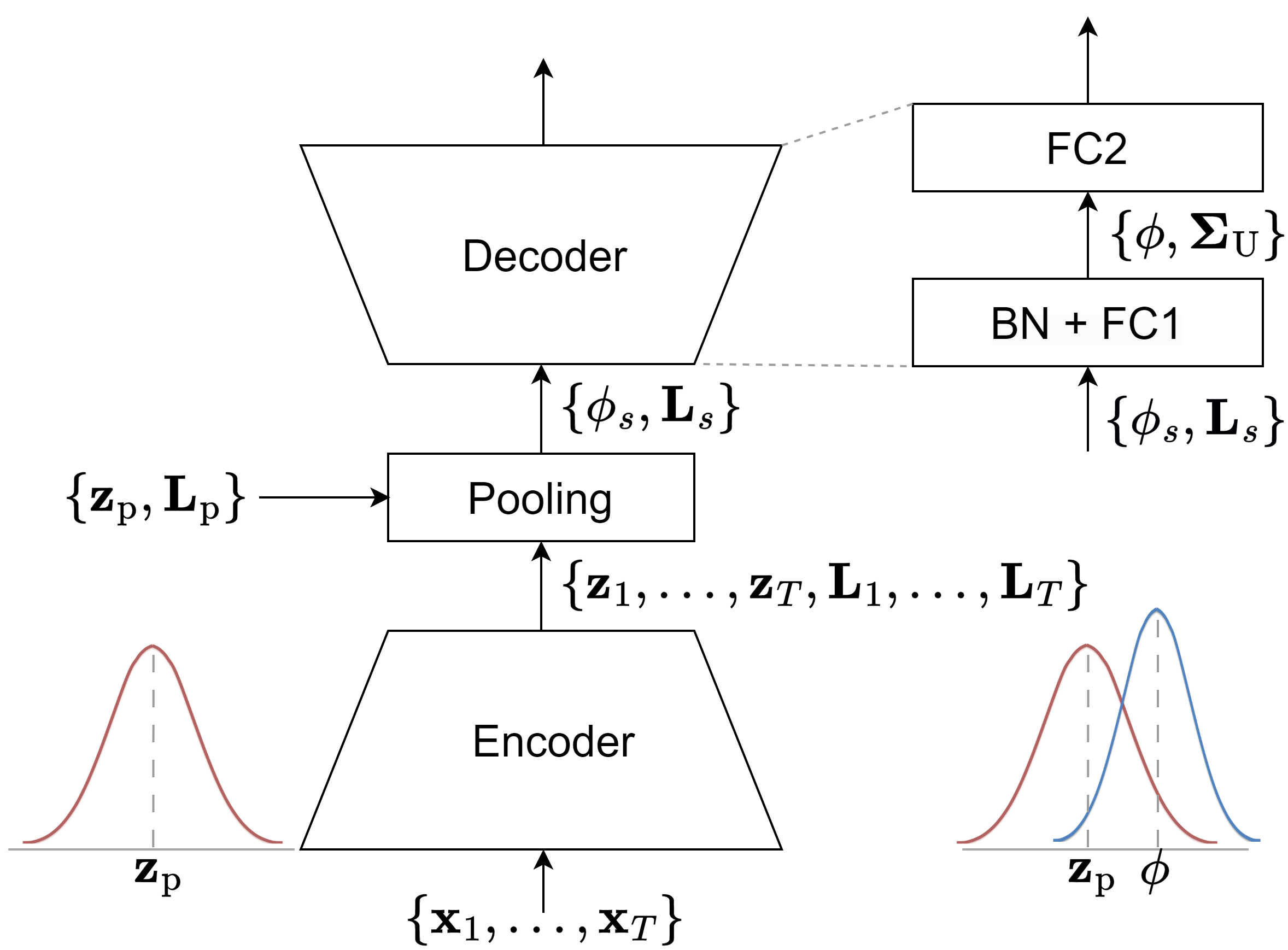}
\vspace{0.1cm}
\caption{\it A speaker-embedding neural network with uncertainty propagation. ``BN'' and ``FC'' are shorthand for batch normalization and a fully-connected layer, respectively. The Gaussian distributions are to illustrate that the posterior distribution of an embedding differs from the prior distribution.}
\label{fig:network}
\vspace{-0.25cm}
\end{figure}
\vspace{-0.2cm}
\subsection{Uncertainty of neural speaker embedding}
\label{ssec:derive}
As shown in Fig.~\ref{fig:network}, we assume that an encoder maps an input sequence $\{\mathbf{x}_1, ..., \mathbf{x}_T\}$ of $T$ frames to another sequence $\{\mathbf{z}_1, ..., \mathbf{z}_T\}$ with precision $\{\mathbf{L}_1^{-1}, ... \mathbf{L}_T^{-1}\}$. The encoder is commonly known as the backbone neural network. Popular architectures are ResNet34~\cite{He16}, ECAPA-TDNN~\cite{Desplanques20}, TDNN~\cite{Snyder18}, and so on.  
By defining the temporal pooling as a Gaussian inference process
~\cite{lee21}, a point estimate 
$\bm{\phi}_s$ and its precision
$\mathbf{L}_s$ are obtained as
\begin{equation}
    \bm{\phi}_s = \mathbf{L}_s ^{-1} \left(\sum_{t=1}^{T} \mathbf{L}_t \mathbf{z}_t + \mathbf{L}_\text{p} \mathbf{z}_\text{p}\right)
    \label{eq:posterior_mean}
\end{equation}
\vspace{-0.2cm}
\begin{equation}
    \mathbf{L}_s = \sum_{t=1}^{T} \mathbf{L}_t + \mathbf{L}_\text{p}
    \label{eq:posterior_prec}
\end{equation}
where $\bm{\mu}_\text{p}$ and $\mathbf{L}_\text{p}$ represent the prior mean and precision, respectively. The posterior precision $\mathbf{L}_s$ measures the uncertainty associated with the point estimate in \eqref{eq:posterior_mean}. Referring to the black arrows in Fig.~\ref{fig:network}, the processes are summarized as follows: 
\begin{equation}
    \{\mathbf{z}_1, ..., \mathbf{z}_T, \mathbf{L}_1^{-1}, ... \mathbf{L}_T^{-1} \} \leftarrow f_\text{enc}\left(\mathbf{x}_1, ..., \mathbf{x}_T\right)
    \label{eq:enc}
\end{equation}
\begin{equation}
    \{ \bm{\phi}_s, \mathbf{L}_s \} \leftarrow g_\text{pool}\left(\mathbf{z}_1, ..., \mathbf{z}_T, \mathbf{L}_1^{-1}, ... \mathbf{L}_T^{-1}\right)
    \label{eq:pool}
\end{equation}

While the posterior precision $\mathbf{L}_s$ is not directly used in the network training~\cite{lee21}, it can be propagated to the speaker-embedding space through the feed-forward layers in a similar manner as the posterior mean (see Fig.~\ref{fig:network}), as follows:
\begin{equation}
    \mathbf{\Sigma}_\text{U} \leftarrow \mathbf{W}\mathbf{\Sigma}_{s,\text{BN}}\mathbf{W}\trans
    \label{eq:postcov}
\end{equation}
\begin{equation}
    \bm{\phi} \leftarrow \mathbf{W}\bm{\phi}_{s,\text{BN}}+\mathbf{b}
    \label{eq:postmean}
\end{equation}
where $\mathbf{W}$ denotes the weight, and $\mathbf{b}$ represents the bias of the first FC layer. The results of batch normalization of $\{\bm{\phi}_{s}, \mathbf{L}_s^{-1}\}$ are represented as  $\{\bm{\phi}_{s,\text{BN}}, \mathbf{\Sigma}_{s,\text{BN}}\}$. More details can be found in~\cite{qwang23}. 
The network outputs the speaker embeddings represented by a distribution with the mean $\bm{\phi}$ and a variance $\mathbf{\Sigma}_{\text{U}}$ that is referred to as the uncertainty of an embedding.

\begin{figure}[t]
\centering
\includegraphics[width=0.75\columnwidth]{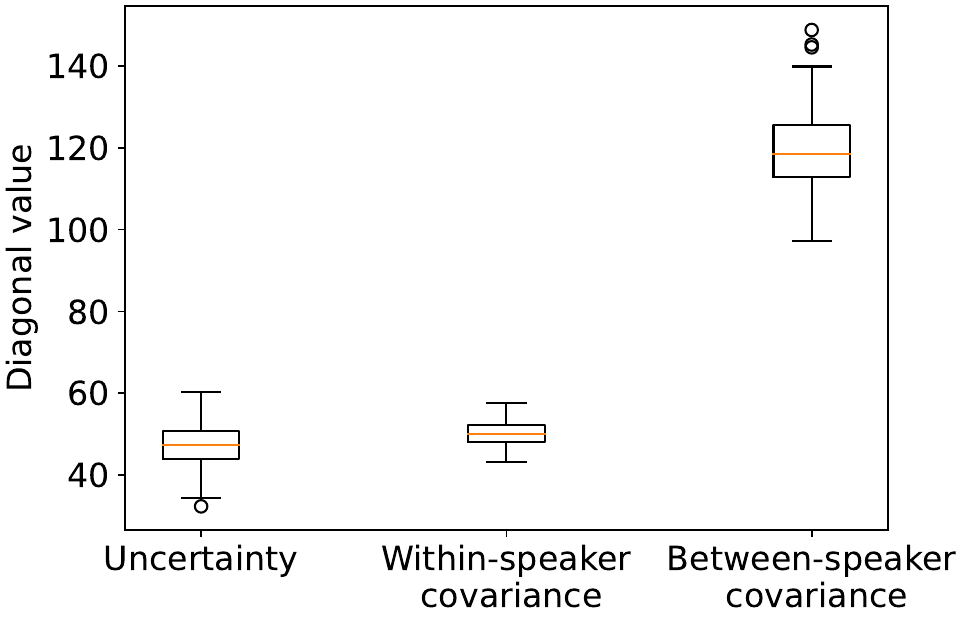}
\caption{\it Boxplots of the diagonal values of the average uncertainty estimation, and the within- and between-speaker covariance matrices estimated on the VoxCeleb2 development set. The off-diagonal values are disregarded.}
\label{fig:cov}
\vspace{-0.25cm}
\end{figure}

To understand the scale of the embedding uncertainty $\mathbf{\Sigma}_{\text{U}}$, we plot the diagonal values of the average uncertainty in Fig.~\ref{fig:cov} alongside that of between- and within-speaker covariance matrices computed from the embedding means.  The off-diagonal values are discarded. The embeddings are extracted from the VoxCeleb2 development dataset~\cite{Chung18} using the xi-vector network~\cite{lee21}. Notably, the between-speaker covariance exhibits significantly higher values than the within-speaker covariance, indicating discriminativeness across speakers and compactness within the same speaker.   This aligns with our findings in~\cite{wang22}. The notable overlap between uncertainty and within-speaker covariance values suggests considering uncertainty in computing within-speaker covariance~\cite{qwang23}. This highlights the need to propagate uncertainty for improving accuracy in back-end scoring.
\vspace{-0.25cm}

\subsection{Cosine scoring}
The objective of a speaker verification task is to verify whether the testing speech is from the enrolled speaker. To achieve this, a similarity is computed between the embeddings of the enrollment speech ($\phi_\text{e}$) and the test speech ($\phi_\text{t}$). One widely used approach is the cosine similarity scoring, known for its computational efficiency. It computes the cosine of the angle between the two embedding vectors, functioning as an indicator of their similarity~\cite{Liu19, Zhou20, Desplanques20, wang22}, as follows
\begin{equation}
s(\phi_\text{e},\phi_\text{t}) = \frac
    {<\phi_\text{e},\phi_\text{t}>}
    {\norm{\phi_\text{e}}\norm{\phi_\text{t}}}
\label{eq:cos1}
\end{equation}
where $<\cdot>$ represents the inner product between two vectors, and $\norm{\cdot}$ signifies $l2$ normalization. Note that cosine scoring in (\ref{eq:cos1}) has no ability to handle uncertainty.

\vspace{-0.25cm}
\section{Uncertainty-aware cosine scoring}
\label{sec:pagestyle}

Cosine scoring (\ref{eq:cos1}) measures the similarity between embeddings on a unit circle. It calculates the Euclidean lengths of the enrollment and test embeddings in the denominator, treating all dimensions equally, as follows: 

\begin{equation}
s(\phi_\text{e},\phi_\text{t}) = \frac
    {\phi\trans_\text{e}\phi_\text{t}}
    {
        \sqrt{\phi\trans_\text{e}\mathbf{I}\phi_\text{e}}
        \sqrt{\phi\trans_\text{t}\mathbf{I}\phi_\text{t}}
    }
\label{eq:cos2}
\end{equation}
With the consideration of uncertainty in the embedding representations, it's evident that each dimension of the embeddings is no more independent or identically distributed. Therefore, they should be treated differently with the reference of covariance $\mathbf{\Sigma}_\text{U}$ in (\ref{eq:postcov}). For example, the same distance in the dimensions with larger variances should be counted as less contribution in the ``length calculation" process, and the opposite for that in the dimensions with smaller variances. 

As uncertainty is additive, the uncertainty posterior covariance of the embedding can be added to the identity matrix with a factor of  $\frac{1}{d}$, as follows:
\begin{equation}
s_\text{UP}= \frac
    {\phi\trans_\text{e}\phi_\text{t}}
    {
        \sqrt{\phi\trans_\text{e}\left(\mathbf{I}+\sigmaue\right)^{-1}\phi_\text{e}}
        \sqrt{\phi\trans_\text{t}\left(\mathbf{I}+\sigmaut\right)^{-1}\phi_\text{t}}
    }
\label{eq:upcos}
\end{equation}
Dividing the uncertainty covariance $\{\mathbf{\Sigma}_\text{Ue}, \mathbf{\Sigma}_\text{Ut} \}$ by the dimensionality of the embedding vectors $d$ keeps the scale of the denominator the same. The uncertainty-aware cosine scoring propagates the uncertainty of speaker embedding extraction to the back-end scoring. 

The uncertainty-propagated cosine (UP-Cos) scoring in (\ref{eq:upcos}) can be interpreted as an inner production between the enrollment and test embeddings normalized by the Mahalanobis distance lengths of the two embeddings.  Mahalanobis distance~\cite{Mahalanobis36} is a multivariate statistical distance measure that scales the contribution of individual variables to the distance value according to their variance. We define the Mahalanobis distance length $\norm{\bm{\phi}}_\text{ML}$ of an embedding as the Mahalanobis distance between the coordinate origin and the endpoint of the embedding vector, as follows:
\begin{equation}
\norm{\bm{\phi}}_\text{ML} = \sqrt{ \left(\bm{\phi} -\bm{0} \right)\trans \mathbf{\Sigma} ^{-1} \left(\bm{\phi}-\bm{0}\right)}
\label{eq:ml_len}
\end{equation}
In the uncertainty-aware cosine scoring (\ref{eq:upcos}), the variance $\mathbf{\Sigma}$ for Mahalanobis distance is 
\begin{align}
\mathbf{\Sigma}_\text{e} = \mathbf{I} + \frac{1}{d}\mathbf{\Sigma}_\text{Ue} &&
\mathbf{\Sigma}_\text{t} = \mathbf{I} + \frac{1}{d}\mathbf{\Sigma}_\text{Ut}
\label{eq:sigmaet}
\end{align}
respectively. 
The scoring can be re-written as 
\begin{equation}
s_\text{UP} = \frac
    {<\bm{\phi}_\text{e},\bm{\phi}_\text{t}>}
    {
        \sqrt{ \bm{\phi}_\text{e} \trans \Sigma_\text{e} ^{-1} \bm{\phi}_\text{e}}
        \sqrt{ \bm{\phi}_\text{t} \trans \Sigma_\text{t} ^{-1} \bm{\phi}_\text{t}}
    }
\label{eq:up-cos}
\end{equation}
In the case where the uncertainty is not considered, i.e., $\mathbf{\Sigma}_\text{U}=0$ and $\Sigma_\text{e}=\Sigma_\text{t}=\mathbf{I}$, the Mahalanobis length (\ref{eq:ml_len}) reduces to the Euclidean length.
In addition, the uncertainty-aware cosine similarity score $s_\text{UP}$ in (\ref{eq:upcos}) becomes equivalent to the original cosine similarity score as in (\ref{eq:cos1}).

Total covariance captures the global covariance of embeddings and can be an indicator for the contribution of each dimension to the Mahalanobis distance. When considering embedding uncertainty, the total covariance attributed to an individual embedding is computed as a sum of the uncertainty matrix and the total covariance $\mathbf{\Sigma}_\text{tot}$ derived from embedding means (see UP-Cos 2 in Table~\ref{tab:tab1}). UP-Cos 3 and UP-Cos 4 are the variations of the previous UP-Cos scorings considering the possible correlation between the enroll and test embeddings.

\begin{figure}[t]
\centering
\includegraphics[width=0.70\columnwidth]{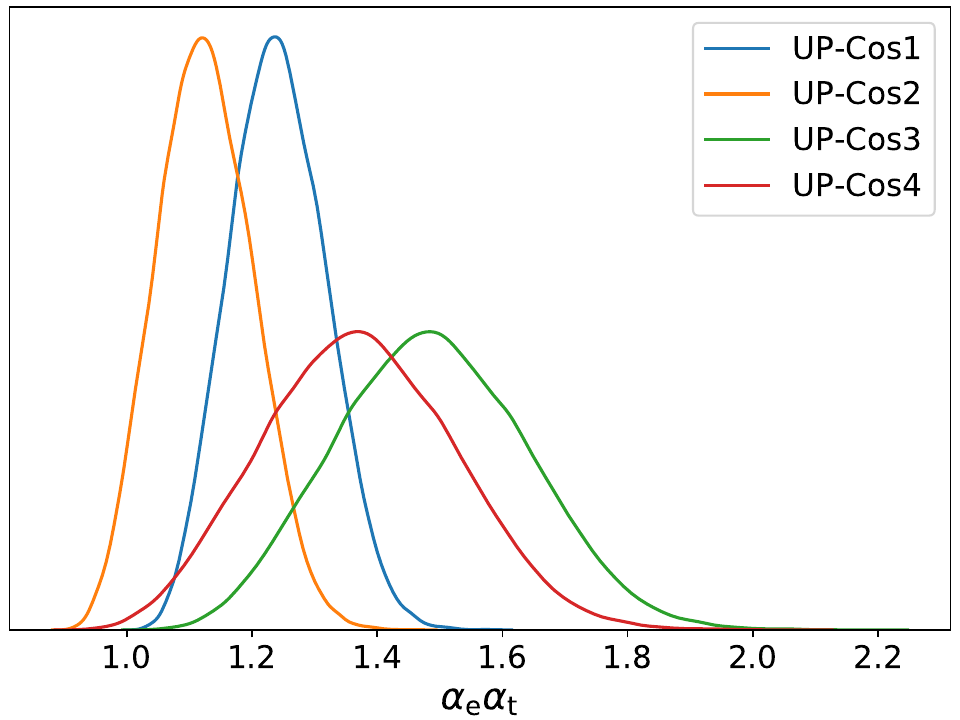} 
\caption{{\it Distribution of the product $\alpha_\text{e} \alpha_\text{t}$ when using the four methods of UP-Cos shown in Table~\ref{tab:tab1}. The investigation is done on the VoxCeleb1-O test dataset.}}
\label{fig:alpha_hist}
\vspace{-0.25cm}
\end{figure}

The relation between the uncertainty-aware cosine similarity and the original cosine similarity is shown as follows:
\begin{equation}
    s_\text{UP}\left( \bm{\phi}_\text{e}, \bm{\phi}_\text{t}\right)= \alpha_\text{e} \alpha_\text{t} \cos\left( \bm{\phi}_\text{e}, \bm{\phi}_\text{t}\right)
    \label{eq:up-cos1}
\end{equation}
where $\alpha_\text{e}$ and $\alpha_\text{t}$ are
\begin{align}
   \alpha_\text{e} 
    = \left(\frac
        {\bm{\phi}_\text{e} \trans \bm{\phi}_\text{e} }
        {\bm{\phi}_\text{e} \trans \Sigma_\text{e} ^{-1} \bm{\phi}_\text{e}}
    \right) ^\frac{1}{2}
    && 
    \alpha_\text{t} 
    = \left(\frac
        {\bm{\phi}_\text{t} \trans \bm{\phi}_\text{t} }
        {\bm{\phi}_\text{t} \trans \Sigma_\text{t} ^{-1} \bm{\phi}_\text{t}}
    \right) ^\frac{1}{2}
    \label{eq:alpha}
\end{align}
Therefore, the uncertainty of enrollment and test speaker embedding representations is reflected in the effect of scaling in the scores. 
The scaling factors $\alpha_\text{e}$ and $\alpha_\text{t} $ are scalars calculated from a function of the embedding vector $\bm{\phi}$ and uncertainty covariance $\mathbf{\Sigma}_\text{U}$.
It is worth noting that the uncertainty influences the range of UP-Cos scores $s_\text{UP}$. Due to the factor $\alpha_\text{e}\alpha_\text{t}$ shown in Fig.~\ref{fig:alpha_hist}, it no longer falls within the range of $\left[-1,1\right]$ as with cosine similarity scores (\ref{eq:cos1}). 

\begin{table}[t]
\begin{center}
\caption{ Four UP-Cos methods}
\begin{tabular}{c | c |c} 
\hline 
Scoring & Enrol $\mathbf{\Sigma}$ & Test $\mathbf{\Sigma}$ \\
\hline \hline 
UP-Cos 1 
& $d^{-1}\mathbf{\Sigma}_{\text{Ue}} + \mathbf{I}$
& $ d^{-1}\mathbf{\Sigma}_{\text{Ut}} + \mathbf{I} $  \\
\hline 
UP-Cos 2 
& $  d^{-1}\left(\mathbf{\Sigma}_{\text{Ue}} + \mathbf{\Sigma}_\text{tot}\right)$
& $  d^{-1}\left(\mathbf{\Sigma}_{\text{Ut}}+\mathbf{\Sigma}_\text{tot}\right ) $ \\
\hline 
UP-Cos 3
& \multicolumn{2}{|c}{$ d^{-1}\left(\mathbf{\Sigma}_{\text{Ue}}+\mathbf{\Sigma}_{\text{Ut}}\right)+ \mathbf{I}$ } \\
\hline 
UP-Cos 4
& \multicolumn{2}{|c}{$ d^{-1}\left(\mathbf{\Sigma}_{\text{Ue}}+\mathbf{\Sigma}_{\text{Ut}}+\mathbf{\Sigma}_\text{tot}\right)$ } \\
\hline 
\end{tabular}
\label{tab:tab1}
\end{center}
\end{table}

\section{Experiments}
\label{sec:exp}
\subsection{Experimental settings}
\label{ssec:exp_set}
We evaluate the proposed methods on both the VoxCeleb1~\cite{Nagrani17} and the Speaker in the Wild (SITW) core-core \textit{eval} sets~\cite{McLaren16}. For the former, we test two trial lists: the original test set Vox1-O and the hard test set Vox1-H. We use the development set of the VoxCeleb2 dataset~\cite{Chung18} as the training data. There are no speakers common to both our training and evaluation sets. Data augmentation is utilized to increase training data diversity and quantity. Techniques involve speed perturbation~\cite{Ko15}, random drops of audio chunks and frequency bands~\cite{Park19}, environmental corruptions using room impulse responses, and adding noise~\cite{Ko17}.

For the front-end, we extract 80-dimensional filterbank features from the utterance audio and input them to speaker-embedding networks. We employ the ECAPA-TDNN backbone with AAM-Softmax cross-entropy loss~\cite{Desplanques20}. Attentive statistics pooling and posterior inference pooling~\cite{lee21} are used in the networks to produce the two types of embedding vectors differing in whether to take uncertainty under consideration in embedding extraction, respectively, referred to as Emb and UP-Emb in the experiments. 
For UP-Emb, frame-wise precision $\mathbf{L}_t$ is assumed to be diagonal~\cite{lee21}, resulting in embeddings' posterior covariance $\mathbf{\Sigma}_{\text{U}}$ having significantly larger diagonal values than off-diagonal ones. We utilize them as diagonal matrices, discarding off-diagonal values.
We use the SpeechBrain open-source toolkit~\cite{sb21} for the implementation of the front-end and extraction of embeddings.

The two baselines are the cosine scoring (Cos) on both types of embeddings. For the proposed scoring with uncertainty (UP-Cos), we use the UP-Emb and propagate the frame-wise uncertainty to the embedding space following~\ref{ssec:derive}. We evaluate four variations of UP-Cos in Table~\ref{tab:tab1}. For UP-Cos 3 and 4, the total covariance is pre-computed using the same training data and subsequently diagonalized for convenience and alignment with uncertainty covariance matrices $\mathbf{\Sigma}_\text{U}$. Additionally, we employ PLDA and uncertainty-propagated PLDA (UP-PLDA)~\cite{qwang23} for comparison. Length scaling (LS) and uncertainty-propagated LS (UP-LS) pre-processing techniques are applied to PLDA and UP-PLDA, respectively, as recommended in~\cite{qwang23}. The training is conducted on the same training data with concatenations of segments belonging to the same session. 
\vspace{-0.25cm}

\subsection{Results and analysis}
\begin{table}[t]
\centering
\caption{Performance of the proposed uncertainty-propagated scoring and baselines on Vox1-O, Vox1-H, and SITW datasets. Emb and UP-Emb represent the embeddings extracted from the ECAPA-TDNN network with attentive statistics pooling and posterior inference pooling, respectively. Four UP-Cos scoring methods are evaluated with UP-Emb. Results are shown as EER/minDCF.}
\begin{tabular}{ c | c c c }
    \hline 
     System & Vox1-O & Vox1-H  & SITW  \\
    \hline \hline 
     Emb+Cos  & 1.22/0.176 & 2.46/0.235 & 1.79/0.170 \\  
     Emb+PLDA & 1.23/0.173 & 2.46/0.236 & 1.69/0.172 \\ 
    \hline 
     UP-Emb+Cos  
    &  1.23/0.145 & 2.42/0.234 & 1.75/0.168 \\
     UP-Emb+PLDA  & 1.23/0.141  & 2.43/0.235 &  1.77/0.168 \\
     UP-Emb+UP-Cos 1  & 1.05/\textbf{0.115} & 2.20/\textbf{0.219} & 1.72/\textbf{0.164} \\
     UP-Emb+UP-Cos 2  & 1.04/0.117 & 2.20/0.221 & 1.72/\textbf{0.164}\\
     UP-Emb+UP-Cos 3  & 1.01/0.127 & \textbf{2.17}/0.233  & 1.67/\textbf{0.164} \\ 
     UP-Emb+UP-Cos 4  & \textbf{0.99}/0.131 & \textbf{2.17}/0.240 & 1.64/0.172 \\
     UP-Emb+UP-PLDA  & 1.01/0.124 & 2.18/0.224 & \textbf{1.59}/0.166 \\
    \hline
   \end{tabular}
    \label{tab:tab2}
    \vspace{-0.1cm}
\end{table}
\begin{figure}[t]
     \centering
     \begin{subfigure}[b]{0.240\textwidth}
         \centering
         \includegraphics[width=\textwidth  ]{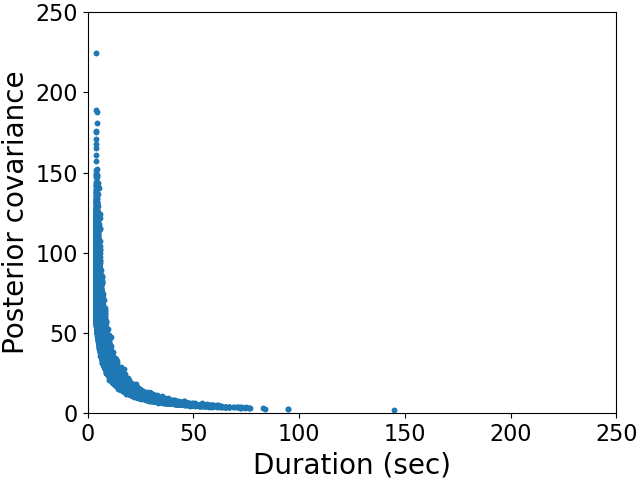}
         \caption{Vox1-H}
         \label{fig:postcov_vox1h}
     \end{subfigure}
     \hfill
     \begin{subfigure}[b]{0.24\textwidth}
         \centering
         \includegraphics[width=\textwidth]{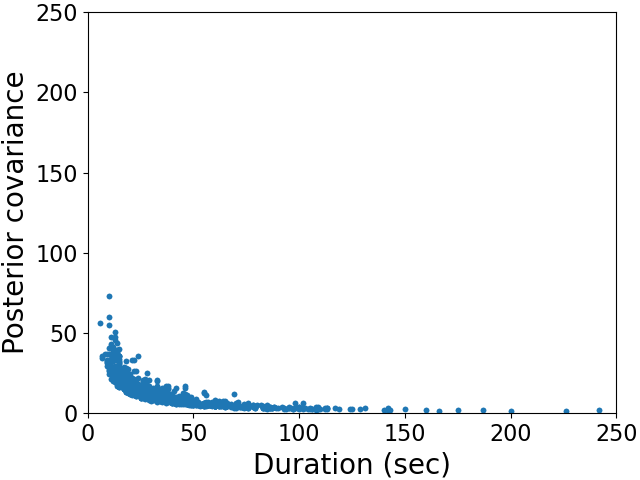}
         \caption{SITW-eval}
         \label{fig:fpostcov_sitw}
     \end{subfigure}
        \caption{\it Scatter plot of average posterior covariance versus utterance duration for Vox1-H and SITW-eval datasets.}
        \label{fig:correlation}
        \vspace{-0.53cm}
\end{figure}

We first compare the four systems of two back-ends (Cos and PLDA) with the conventional embeddings (Emb) and uncertainty-aware embeddings (UP-Emb), as shown in the first four rows in Table~\ref{tab:tab2}. We observe that with the Emb, the use of Cos and PLDA give similar performance. It is expected since a large margin softmax is used in the network training~\cite{wang22}. 
In both Cos and PLDA systems, UP-Emb consistently shows superior or comparable performance in comparison to Emb. 
It is confirmed that considering uncertainty in deriving speaker embedding extraction helps extract better speaker embeddings for speaker recognition tasks, while the effect is limited.

We next propagate the uncertainty, represented as posterior covariance, to the speaker embedding space and evaluate the proposed cosine scoring with uncertainty. 
As shown in Table~\ref{tab:tab2}, all of the UP-Cos methods outperform the Cos baseline in terms of EER over all three evaluation sets, and UP-Cos 1, 2, and 3 reduce minDCF in addition. It indicates that further propagating the uncertainty to the back-end is helpful for speaker recognition tasks, and the proposed cosine scoring with uncertainty has the capability to handle the uncertainty from embedding estimation in cosine similarity. 
According to Table~\ref{tab:tab1}, $\mathbf{\Sigma}$ for UP-Cos 1 and 2 are dependent on the embedding only, while for UP-Cos 3 and 4, $\mathbf{\Sigma}$  depends on both the enrollment and test embeddings in a trial. Therefore, UP-Cos 1 and 2 are more computationally efficient in the testing phase. In addition, the total covariance in UP-Cos 2 and 4 is calculated, in advance, from a dataset other than the evaluation set. Considering the computational cost, UP-Cos 1 is the most recommended for practical use. It achieved consistent improvement with average 8.5\% and 9.8\% reductions in EER and minDCF, respectively compared with the baseline.

We further compare the use of uncertainty in cosine scoring and PLDA scoring in Table.~\ref{tab:tab2}. We observe a consistency with~\cite{qwang23} that UP-PLDA outperforms PLDA when using UP-Emb. Compared with UP-Cos 1, UP-PLDA achieves better EERs but worse minDCF in Vox1 sets while the opposite in the SITW set. Compared to UP-PLDA, UP-Cos takes much less computational cost. 

Finally, we investigate the correlation between uncertainty with speech duration. By averaging the diagonal values of each posterior covariance, across the 192 dimensions, we generate scalars.  They are then plotted against the utterance duration in Fig.~\ref{fig:correlation}. A negative correlation is observed between the two variables. It is consistent with our expectation that embeddings extracted from a shorter utterance are less reliable, i.e., larger uncertainty. Therefore, it is confirmed that the posterior covariance measures the uncertainty, and is related to its nature. The posterior covariance of SITW-eval exhibits higher values compared to that of the Vox-1h for the same speech duration, possibly indicating a domain mismatch from the training data. Consequently, the extracted embeddings show increased uncertainties.

\section{Conclusion}
We have shown the importance of taking uncertainty into consideration in speaker recognition because of speaker-independent variabilities exhibited in speech utterances and the extraction networks. On the basis of neural speaker embeddings, we show how to propagate the posterior covariance (i.e., uncertainty) to the back-end. We also propose uncertainty-aware scoring. To the best of our knowledge, it is the first work of non-parametric scoring with uncertainty for speaker recognition. Experiments confirmed the efficacy of the proposed scoring and showed consistent improvement in performance on the VoxCeleb and the SITW datasets.  It achieved consistent improvement with 9.5\%/15.0\% average reductions in EER/minDCF compared to the setting in which no uncertainty is considered and is 8.5\%/9.8\% compared to the cosine similarity following certainty-aware embeddings. 

\bibliographystyle{IEEEtran}

\bibliography{refs}

\end{document}